\documentclass[12pt]{article}
\usepackage{epsfig}
\usepackage{amsmath}
\usepackage{hhline}
\usepackage{amssymb}
\usepackage{times}
\usepackage{cite}

\newlength{\dinwidth}
\newlength{\dinmargin}
\begin{document}  
\newcommand{\pom}{{I\!\!P}}
\newcommand{\reg}{{I\!\!R}}
\newcommand{\gap}{\stackrel{>}{\sim}}
\newcommand{\lap}{\stackrel{<}{\sim}}

\newcommand{\alps}{\alpha_s}
\newcommand{\sqrts}{$\sqrt{s}$}
\newcommand{\LO}{$O(\alpha_s^0)$}
\newcommand{\Oa}{$O(\alpha_s)$}
\newcommand{\Oaa}{$O(\alpha_s^2)$}
\newcommand{\PT}{p_{\perp}}
\newcommand{\PO}{I\!\!P}
\newcommand{\xpomlo}{3\times10^{-4}}  
\newcommand{\dgr}{^\circ}
\newcommand{\pbarnt}{\,\mbox{{\rm pb$^{-1}$}}}

%
%
\newcommand{\tg}{\theta_{\gamma}}
\newcommand{\te}{\theta_e}
%
%
\newcommand{\qsq}{\mbox{$Q^2$}}
\newcommand{\Qsq}{\mbox{$Q^2$}}
\newcommand{\s}{\mbox{$s$}}
\newcommand{\ttra}{\mbox{$t$}}
\newcommand{\modt}{\mbox{$|t|$}}
\newcommand{\eminpz}{\mbox{$E-p_z$}}
\newcommand{\eminpzs}{\mbox{$\Sigma(E-p_z)$}}
\newcommand{\rap}{\ensuremath{\eta^*} }
\newcommand{\W}{\mbox{$W$}}
\newcommand{\w}{\mbox{$W$}}
\newcommand{\Q}{\mbox{$Q$}}
\newcommand{\q}{\mbox{$Q$}}
\newcommand{\xB}{\mbox{$x$}}  
\newcommand{\xF}{\mbox{$x_F$}}  
\newcommand{\xg}{\mbox{$x_g$}}  
\newcommand{\xbj}{x}
\newcommand{\xpom}{x_{\PO}}
\newcommand{\y}{\mbox{$y~$}}

%
\newcommand{\gp}{\ensuremath{\gamma^*}p }
\newcommand{\gammasp}{\ensuremath{\gamma}*p }
\newcommand{\gammap}{\ensuremath{\gamma}p }
\newcommand{\gsp}{\ensuremath{\gamma^*}p }
\newcommand{\epem}{\mbox{$e^+e^-$}}
\newcommand{\ep}{\mbox{$ep~$}}
\newcommand{\epl}{\mbox{$e^{+}$}}
\newcommand{\emi}{\mbox{$e^{-}$}}
\newcommand{\epm}{\mbox{$e^{\pm}$}}
%
\newcommand{\photon}{\mbox{$\gamma$}}
\newcommand{\phib}{\mbox{$\varphi$}}
\newcommand{\rh}{\mbox{$\rho$}}
\newcommand{\rhz}{\mbox{$\rh^0$}}
\newcommand{\ph}{\mbox{$\phi$}}
\newcommand{\om}{\mbox{$\omega$}}
\newcommand{\ome}{\mbox{$\omega$}}
\newcommand{\jpsi}{\mbox{$J/\psi$}}
\newcommand{\JPSI}{J/\psi}
\newcommand{\ups}{\mbox{$\Upsilon$}}
\newcommand{\bsl}{\mbox{$b$}}
%
%
\newcommand{\cm}{\mbox{\rm cm}}
\newcommand{\GeV}{\mbox{\rm GeV}}
\newcommand{\gev}{\mbox{\rm GeV}}
\newcommand{\GeVx}{\rm GeV}
\newcommand{\gevx}{\rm GeV}
\newcommand{\MeV}{\mbox{\rm MeV}}
\newcommand{\mev}{\mbox{\rm MeV}}
\newcommand{\MeVx}{\mbox{\rm MeV}}
\newcommand{\mevx}{\mbox{\rm MeV}}
\newcommand{\GeVsq}{\mbox{${\rm GeV}^2$}}
\newcommand{\gevsq}{\mbox{${\rm GeV}^2$}}
\newcommand{\gevsqc}{\mbox{${\rm GeV^2/c^4}$}}
\newcommand{\gevcsq}{\mbox{${\rm GeV/c^2}$}}
\newcommand{\mevcsq}{\mbox{${\rm MeV/c^2}$}}
\newcommand{\GeVsqm}{\mbox{${\rm GeV}^{-2}$}}
\newcommand{\gevsqm}{\mbox{${\rm GeV}^{-2}$}}
\newcommand{\nb}{\mbox{${\rm nb}$}}
\newcommand{\nbinv}{\mbox{${\rm nb^{-1}}$}}
\newcommand{\pbinv}{\mbox{${\rm pb^{-1}}$}}
\newcommand{\mm}{\mbox{$\cdot 10^{-2}$}}
\newcommand{\mmm}{\mbox{$\cdot 10^{-3}$}}
\newcommand{\mmmm}{\mbox{$\cdot 10^{-4}$}}
\newcommand{\degr}{\mbox{$^{\circ}$}}
%
%
\def\gsim{\,\lower.25ex\hbox{$\scriptstyle\sim$}\kern-1.30ex%
  \raise 0.55ex\hbox{$\scriptstyle >$}\,}
\def\lsim{\,\lower.25ex\hbox{$\scriptstyle\sim$}\kern-1.30ex%
  \raise 0.55ex\hbox{$\scriptstyle <$}\,}
%
%

\begin{titlepage}

\noindent
DESY 01--093  \hfill  ISSN 0418-9833 \\
June 2001

\vspace{2cm}

\begin{center}
\begin{Large}

{\bf Measurement of Deeply Virtual Compton Scattering\\
 at HERA}

\vspace{2cm}

H1 Collaboration

\end{Large}
\end{center}

\vspace{2cm}

\begin{abstract}
A measurement is presented of elastic Deeply Virtual Compton 
Scattering $e^+ + p \rightarrow e^+ + \photon + p$ at HERA using 
data taken with the H1 detector.
The cross section is measured as a function of the photon 
virtuality, $Q^2$, and the invariant mass, $W$, of the  
$\gamma p$ system, in the kinematic range 
$2 < Q^2 < 20\,{\rm GeV}^2 $,  $30 < W < 120\,{\rm GeV}$ and 
$|t| < 1\,{\rm GeV}^2$, where $t$ is the squared momentum transfer 
to the proton.
The measurement is compared to QCD based calculations.

\end{abstract}

\vspace{1.5cm}

\begin{center}
To be submitted to Physics Letters B
\end{center}

\end{titlepage}

%
%

\begin{flushleft}

C.~Adloff$^{33}$,              
V.~Andreev$^{24}$,             
B.~Andrieu$^{27}$,             
T.~Anthonis$^{4}$,             
V.~Arkadov$^{35}$,             
A.~Astvatsatourov$^{35}$,      
A.~Babaev$^{23}$,              
J.~B\"ahr$^{35}$,              
P.~Baranov$^{24}$,             
E.~Barrelet$^{28}$
% ...

\bigskip{\it
% ...
$ ^{4}$ Inter-University Institute for High Energies ULB-VUB, Brussels;
          Universitaire Instelling Antwerpen, Wilrijk; Belgium$^{ c}$ \\
% ...
$ ^{23}$ Institute for Theoretical and Experimental Physics,
          Moscow, Russia$^{ l}$ \\
 $ ^{24}$ Lebedev Physical Institute, Moscow, Russia$^{ e,h}$ \\
% ...
$ ^{27}$ LPNHE, Ecole Polytechnique, IN2P3-CNRS, Palaiseau, France \\
 $ ^{28}$ LPNHE, Universit\'{e}s Paris VI and VII, IN2P3-CNRS,
          Paris, France \\
% ...
$ ^{33}$ Fachbereich Physik, Bergische Universit\"at Gesamthochschule
          Wuppertal, Wuppertal, Germany$^{ a}$ \\
% ...
$ ^{35}$ DESY, Zeuthen, Germany$^{ a}$ \\
% ...
$ ^a$ Supported by the Bundesministerium f\"ur Bildung, Wissenschaft,
      Forschung und Technologie, FRG,
      under contract numbers 7AC17P, 7AC47P, 7DO55P, 7HH17I, 7HH27P,
      7HD17P, 7HD27P, 7KI17I, 6MP17I and 7WT87P \\
% ...
$ ^c$ Supported by FNRS-NFWO, IISN-IIKW \\
% ...
$ ^e$ Supported by the Deutsche Forschungsgemeinschaft \\
% ...
$ ^h$ Supported by Russian Foundation for Basic Research
      grant no. 96-02-00019 \\
% ...
$ ^l$ Partially Supported by Russian Foundation
      for Basic Research, grant
    no. 00-15-96584 \\
}

\end{flushleft}

\newpage

\section{Introduction}
\noindent

\begin{figure}[htb]
 \begin{center}
  \epsfig{figure=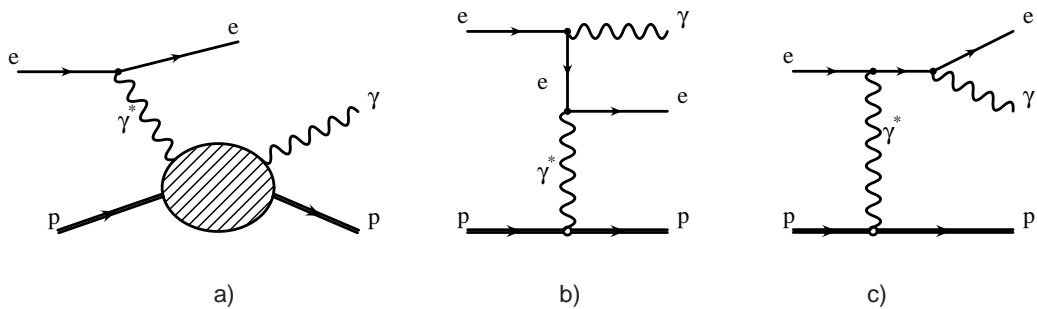,width=0.9\textwidth}
  \caption{\sl The DVCS (a) and the Bethe-Heitler (b and c) processes.}
  \label{fig:bh}
 \end{center}
\end{figure}

This paper presents the first measurement of the elastic 
cross section for Deeply Virtual Compton Scattering (DVCS) 
(Fig.~\ref{fig:bh}a) i.e.~the diffractive scattering of a 
virtual photon off a proton~\cite{Ji:1997nm, Collins:1999be, 
Ji:1998xh, Mankiewicz:1998bk, Belitsky:2000sg, Blumlein:2000cx}
by studying the reaction:
\begin{equation}
e^+ + p \rightarrow e^+ + \photon + p .
\end{equation}
\label{eq:reac}
The interest of this process, because of its apparent simplicity, 
resides in the 
particular insight it gives for the applicability of 
perturbative Quantum Chromo Dynamics (QCD) in the field of 
diffractive interactions.
The wide kinematic range in the photon virtuality, \qsq, 
accessible at HERA, provides a powerful probe for the interplay 
between perturbative and non-perturbative regimes in QCD.
Furthermore the DVCS process gives access
to a new class of parton distribution functions, 
the skewed parton distributions (SPD)~\cite{Muller:1994fv, Ji:1997ek,
Radyushkin:1997ki} which are generalisations of the familiar 
parton distributions and include parton momentum correlations.

The reaction studied receives contributions from both the DVCS 
process whose origin lies in the strong interaction, and the purely 
electromagnetic Bethe-Heitler (BH) process (Fig.~\ref{fig:bh}b and c) 
where the photon is emitted from the positron line. 
The BH process is precisely known as it depends only on QED 
calculations and proton elastic form factors.
The DVCS cross section is extracted by subtracting the BH
contribution from the total cross section, which is possible 
since the interference contribution
vanishes when averaged over the full azimuthal angle 
of the final state particles. 
A recent measurement of the single spin asymmetry in a longitudinally
polarised electron beam~\cite{Airapetian:2001yk} complements this measurement.

\begin{figure}[htb]
 \begin{center}
  \epsfig{figure=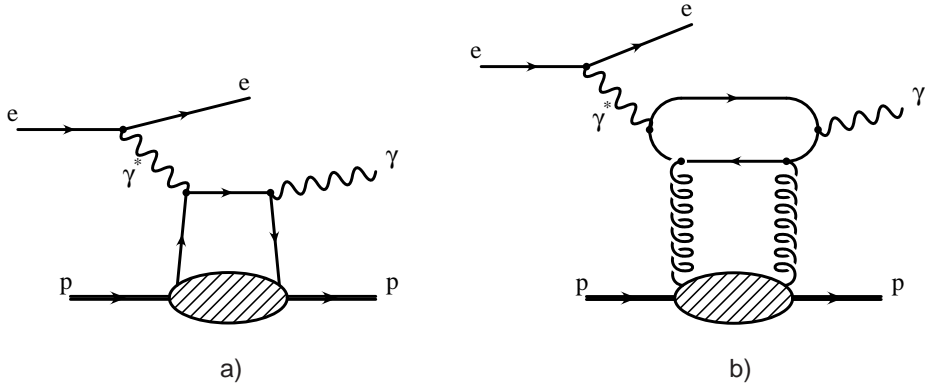,width=0.80\textwidth}
  \caption{\sl The two leading DVCS diagrams in a QCD picture.}
  \label{fig:dvcs}
 \end{center}
\end{figure}

The QCD interpretation of DVCS is based on the two diagrams shown in
Fig.~\ref{fig:dvcs}.
%
%
%
In the presence of a hard scale, the DVCS scattering amplitude
factorises~\cite{Radyushkin:1997ki,Collins:1999be,Ji:1998xh}
into a hard scattering part calculable in perturbative QCD
and parton distributions which contain the non-perturbative 
effects due to the proton structure.
The DVCS process is similar to diffractive vector meson
electro-production, a real photon replacing the final state 
vector meson.
Recent measurements of diffractive vector meson production
at HERA~\cite{h1_rho, zeus_rho_jpsi_hq, h1_jpsi_hq, h1_phi_96, 
zeus_phi_hq} indicate that \qsq\ is relevant for the hard scale 
of the interaction in QCD calculation.
In comparison to vector meson production, DVCS avoids the 
theoretical complications and uncertainties associated with 
the unknown vector meson wave function.
However, even at $Q^2$ values above a few GeV$^2$, non perturbative
effects influence the predictions and have to be modeled.
The DVCS cross section is suppressed relative to that of vector 
meson production by the additional electromagnetic coupling.

Calculations of the DVCS cross section have been published 
by Frankfurt, Freund and Strikman
(FFS)~\cite{Frankfurt:1998at} and by  Donnachie and 
Dosch (DD)~\cite{Donnachie:2000px}.
Both contain ``soft" (non-perturbative) and ``hard" contributions.
The soft part in the FFS prediction is based on the 
aligned jet model~\cite{Bjorken:1973gc}, whereas reggeon and soft 
pomeron exchanges are considered in the DD prediction.
The hard contribution in FFS is calculated in perturbative QCD.
The hard contribution in DD is based on a dipole model where all 
parameters are determined from $pp$ and $\gamma^*p$ cross section 
measurements. 
These predictions only provide the scattering amplitude at 
$t=t_{min}\simeq -m^2_p Q^4/W^4$, where $|t|$ is the squared 
momentum transfer to the proton, $m_p$ the proton mass  
and $W$ the invariant mass of the $\gamma^*p$ system.
In both cases an exponential $t$-dependence, $e^{-b|t|}$, is 
assumed.

For the DVCS process as well as for the vector meson production, 
the transition from a virtual photon to an on shell particle
forces the fractional momenta of the two partons involved to be
unequal and imposes a correlation.
Therefore, the cross section 
calculation necessitates the use of skewed parton 
distributions~\cite{Muller:1994fv, Ji:1997ek, Radyushkin:1997ki}.
The difference in fractional momenta, the skewedness,
becomes important at high \qsq\ values or high vector meson masses.
In particular SPDs have been introduced in order to reconcile QCD 
calculations with the \ups\ diffractive photo-production
measurements at HERA~\cite{zeus_ups_gp,h1_jpsi_gp}.
For vector meson production SPDs appear quadratically in the cross 
section expression.
A unique feature of DVCS is that they appear linearly in the 
interference term with the Bethe-Heitler process.
Therefore the DVCS measurement, in contrast to vector meson production,
offers a particularly suitable channel to extract skewed parton 
distributions~\cite{Diehl:1997bu, Frankfurt:1998at,Freund:2000xf,
Belitsky:2001gz}. 

In this paper, the elastic DVCS cross section measurement at HERA
is presented differentially in $Q^2$ and $W$ in the \qsq\ range 
from $2$ to $20\,$GeV$^2$, \W\ from $30$ to $120\,$GeV and \modt\ below
$1\,$GeV$^2$.

\section{H1 Detector, Kinematics and Monte Carlo Simulation}

A detailed description of the H1 detector can be found in~\cite{h1dect}.
Here only the detector components relevant for the present analysis are
shortly described.
The SPACAL~\cite{spacal},
a lead -- scintillating fibre calorimeter covers the backward\footnote{
H1 uses a right-handed coordinate system with the $z$ axis taken along
the beam direction, the $+z$ or ``forward" direction being that of the 
outgoing proton beam, the ``transverse'' directions are perpendicular 
to the $z$ axis.
The polar angle $\theta$ is defined with respect to the $z$ axis.
The pseudo-rapidity is defined by $\eta=-\ln \tan \theta /2$.}
region of the H1 detector ($ 153 ^{\rm \circ} < \theta < 177.5 ^{\rm
\circ}$).
Its energy resolution for electromagnetic showers is $\sigma(E)/E
\simeq 7.1\%/\sqrt{E/GeV} \oplus 1\%$. 
The uncertainty on the alignment of the calorimeter corresponds to an
uncertainty of $1.3\,$mrad on the scattered positron polar angle.
The liquid argon (LAr) calorimeter ($4^{\rm \circ} \leq \theta \leq
154^{\rm \circ}$) is situated inside a solenoidal magnet. 
The energy resolution for electromagnetic showers is 
$\sigma(E)/E \simeq 11\%/\sqrt{E/GeV}$ as obtained from test beam 
measurements~\cite{Andrieu:1994yn}.
The major components of the central tracking detector are 
two \mbox{2\,m} long coaxial cylindrical drift chambers, 
the CJC, with wires parallel to the beam direction.
The measurement of charged particle transverse momenta is performed
in a magnetic field of $1.15\,$T, uniform over the full tracker volume.
The forward components of the detector, used here to tag hadronic
activity at high pseudo-rapidity ($5 \lsim \eta \lsim 7$), are
the forward muon spectrometer (FMD) and the proton remnant 
tagger (PRT).
The FMD, designed to identify and measure the momentum of muons 
emitted in the forward direction, contains six active layers, each 
made of a pair of planes of drift cells.
The three layers between the main calorimeter and the toroidal
magnet can be reached by secondary particles arising from the
interaction of particles scattered under small angles hitting the 
beam collimators or the beam pipe walls.
Secondary particles or the scattered proton at large \modt\ can 
also be detected by the PRT, located at $24\,$m from the interaction 
point and consisting of double layers of scintillator surrounding 
the beam pipe. 
The trigger used is based on the detection of a cluster in the
electromagnetic section of the SpaCal calorimeter with an energy 
greater than 6~GeV. 

The data were obtained with the H1 detector in the 1997 running period 
when the HERA collider was operated with $820\,$GeV protons and $27.5\,$GeV
positrons.
The sample corresponds to an integrated luminosity of $8\,$\pbinv.

The reconstruction method for the kinematic variables $Q^2$ and 
$x$-Bjorken relies on the polar angle measurements of the final 
state electron, $\te$, and photon, $\tg$, (double angle method):
\begin{eqnarray}
Q^2 & = & 4 E_0^2\, \frac{\sin\ \tg\ (1+\cos \te)}
                {\sin\ \tg\ + \sin\ \te\ - \sin\ (\te\ + \tg)} \\
x & = & \frac{E_0}{E_p} \, \frac{\sin\ \tg\ + \sin\ \te\ + \sin\ (\te\ +
\tg)}
                {\sin\ \tg\ + \sin\ \te\ - \sin\ (\te\ + \tg)} \\
   W^2 & =&  \frac{Q^2}{x}\, (1-x) 
\end{eqnarray}
where $E_0$ and $E_p$ are the electron and the proton beam
energies, respectively.
In case a vertex cannot be reconstructed, the nominal position 
of the $ep$ interactions is taken.
The variable $t$, the square of the four-momentum transfer to the
proton, is very well approximated by the negative square of the 
transverse momentum of the outgoing proton.
The latter is computed as the vector sum of the transverse momenta 
of the final state photon $\vec p_{t_{\gamma}}$ and of the scattered 
positron $\vec p_{t_{e}}$:
\begin{equation}
  t \simeq - |\vec p_{t_{\gamma}} + \vec p_{t_{e}}|^2 \ .
\label{eq:t}
\end{equation}


Monte Carlo simulations are used to estimate the corrections to be
applied to the data for the acceptance and resolutions of the detector. 
The generated events are passed through a detailed simulation of the H1
detector and are subject to the same reconstruction and analysis chain
as the data.
The DVCS process is simulated (for more details see~\cite{rainer})
according to the predicted cross section of FFS~\cite{Frankfurt:1998at},
which includes the DVCS process, the Bethe-Heitler process and 
their interference\footnote{ The free parameters have been set in 
the simulation to the following values: the $t$-slope parameter 
$b=7\,$GeV$^{-2}$, the phase of the QCD amplitude,
$\eta_{QCD}= 1 - \frac{\pi}{2} ( 0.176+0.33 \log{Q^2})$~\cite{eta}
and the sensitivity to the skewedness of the parton densities 
$R_{\gamma}= 0.55$~\cite{Frankfurt:1998at}. 
The proton structure function as extracted from the H1 
data~\cite{h1fit} has been used.}.
Photon radiation from the incoming positron has been included in 
the simulation in the collinear approximation.
The Monte Carlo generator COMPTON 2.0~\cite{compton2} is used to 
simulate Bethe-Heitler events. 
To simulate the background sources (see section 3),  
diffractive \rh,\ \om\ and \ph\ meson events are generated with the 
DIFFVM Monte Carlo~\cite{diffvm}, dilepton production through a 
photon-photon interaction is simulated using the GRAPE 
program~\cite{grape}. 

\section{Event Selection}  \label{sect:selection}

The cross section of the Bethe-Heitler process, proceeding via 
Bremsstrahlung from the positron lines, is the largest 
when the positron and the photon are both produced in the backward 
direction.
In the DVCS case, the final state photon 
does not originate from the
positron and therefore the ratio
of DVCS over BH process is expected to increase when the photon is 
found in the forward direction. 
The analysis is thus restricted to the case where the photon is 
detected in the central or in the forward parts of the detector, 
i.e.~in the LAr calorimeter. 
A data sample, dominated by Bethe-Heitler events, is used as a 
reference sample to monitor the detector performance and its 
simulation.
Two event samples are selected.
\begin{itemize}
\item {\bf Enriched DVCS sample:}
The photon candidate is detected in the  LAr calorimeter and the 
positron candidate in the SpaCal calorimeter.
Both DVCS and Bethe-Heitler processes are expected to contribute to 
this sample with similar magnitudes.
\item {\bf Control sample:}
 The photon candidate is detected in the SpaCal calorimeter and
the positron candidate in the LAr calorimeter.
The contribution of DVCS to this sample is negligible.
\end{itemize}

The event selection is based on the detection of exactly two 
electromagnetic clusters, corresponding to the final state photon and
positron.  One cluster is required to be detected in the SpaCal 
calorimeter with energy larger than $15\,{\rm GeV}$
and the other one in the LAr calorimeter ($25^{\circ} - 145^\circ$) 
with a transverse momentum $p_t > 1\,{\rm GeV}$.
Events with more than one track are rejected. Events with one track 
are only kept if the track is associated to one of the clusters which 
hence identifies the positron candidate.
If no track is reconstructed, the SpaCal cluster is assumed
to be the positron.
In order to reject inelastic and proton dissociation events, 
no further cluster in the LAr calorimeter with energy above 
$0.5\,{\rm GeV}$ is allowed and the absence of activity above 
the noise level in forward detectors PRT and FMD is required. 
The influence of QED radiative corrections is reduced by requirements on 
the longitudinal momentum balance\footnote{ The quantity $\sum E - P_z$ 
is required to be above $45\,{\rm GeV}$. $E$ denotes the energy and  
$P_z$ is the momentum along the beam axis of the final state 
particles. 
The sum is calculated for the final 
state positron and photon.}.
To enhance the DVCS signal with respect to the Bethe-Heitler 
contribution, and to maintain a large detector acceptance, the 
kinematic domain is explicitly restricted to:
$2 < Q^2 < 20\,{\rm GeV}^2 $, $ |t| < 1\,{\rm GeV}^2$ and
$30 < W < 120\,{\rm GeV}$. 
It has to be noted that for the BH process, the $Q^2$ and $W$ variables
cannot be associated with the photon virtuality and the $\gamma^*p$
center of mass energy, respectively.

\subsection{Control sample} \label{sect:sel_bh}

\begin{figure}
  \psfig{figure=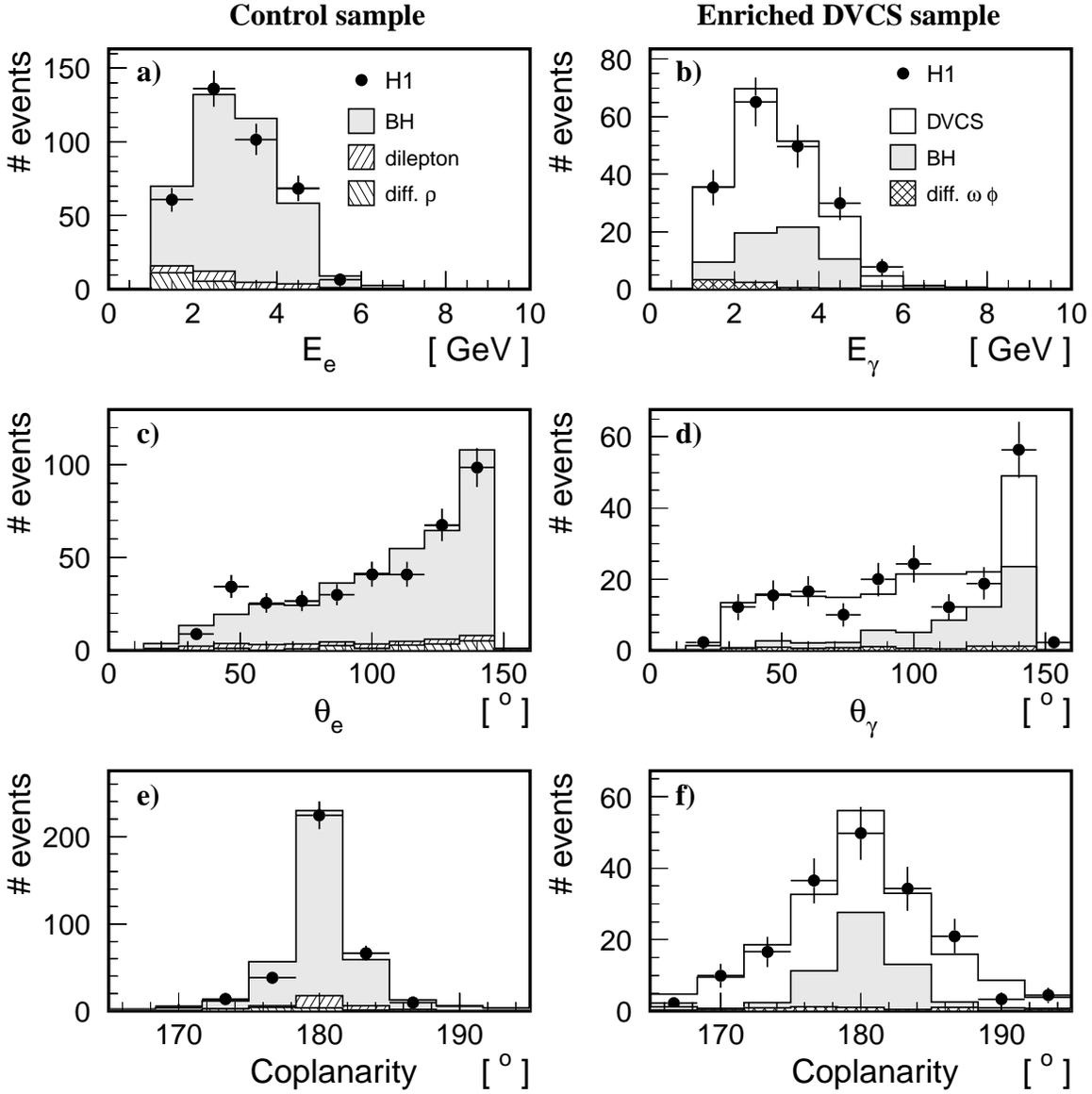,width=17.0cm}
  \caption{\sl Event distributions of the control sample (left) 
and of the enriched DVCS sample (right). 
a-b) energy of the cluster in the LAr calorimeter, 
c-d) polar angle of the cluster in the LAr calorimeter, 
e-f) coplanarity, i.e.~difference of the azimuthal angle of 
the positron and photon candidates. 
The error bars on data points are statistical.
{\it Control sample}: the cluster in the LAr calorimeter corresponds to
the positron candidate. 
The data are compared to the sum of the predictions for the 
Bethe-Heitler process, elastic dilepton production and diffractive 
\rh\ production.
All predictions are normalised to luminosity.
{\it Enriched DVCS sample}: the cluster in the LAr calorimeter 
corresponds to the photon candidate. 
 The data are compared to the sum of the predictions for the 
$e^+ p \rightarrow e^+ \photon p$ reaction according to FFS, 
added to \om\ and \ph\ diffractive backgrounds.
The backgrounds and the BH contribution (shown on top of the backgrounds) 
are normalised to luminosity whereas the DVCS prediction is 
normalised in such a way that the sum of all contributions is equal 
to the total number of events.
}
 \label{fig:cont}
\begin{picture}(0,0)(0,4)
\put(34,230.0){\bf Control sample}
\put(100,230.0){\bf Enriched DVCS sample}
\put(21,222.0){\bf a)}
\put(97,222.0){\bf b)}
\put(21,171.0){\bf c)}
\put(97,171.0){\bf d)}
\put(21,120.0){\bf e)}
\put(97,120.0){\bf f)}
\end{picture}
\end{figure}

This sample of 338 events is dominated by the Bethe-Heitler process. 
Due to the large scattering angle of the positron, the DVCS process is 
suppressed to negligible levels.
In order to have a control of the detector response in the same energy
and angular ranges as for the enriched DVCS sample, the kinematic cuts 
in \qsq\ and $W$ are applied to this sample, treating the photon candidate 
in SpaCal as the scattered positron and the positron candidate in the 
LAr calorimeter as the photon.
Background contributions from inelastic Bethe-Heitler events, diffractive 
$\rho$ meson production and electron pair production have to be 
considered.
The contribution of inelastic Bethe-Heitler events is estimated 
to be $7.7\pm 3.8\,\%$.
The diffractive electro production of \rh\ mesons constitutes a 
background when one of the decay pions is mis-identified as a 
positron in the calorimeter and the other pion is undetected,
while the positron scattered into the SpaCal calorimeter is taken 
to be the photon.
The elastic production of electron pairs by photon-photon processes
$e^+ p \rightarrow e^+ e^- e^+ p$ contributes to the background when
only two of the three leptons are detected. 
These processes have been simulated using the Monte Carlo programs
COMPTON 2.0 (both for elastic and inelastic BH), DIFFVM and GRAPE 
respectively.
In the left column of Fig.~\ref{fig:cont}, event distributions of the 
control sample are shown. 
The data are compared to the sum of the absolute MC predictions which 
includes the Bethe-Heitler process, elastic \rh\ production and elastic 
dilepton production, all normalised to luminosity.
A good description of the data by the sum of the different MC samples is 
achieved, showing that the detector response is well described by 
the simulation.

\subsection{Enriched DVCS sample} \label{sect:sel_dvcs}

This sample (172 events) is found to be dominated by the DVCS 
contribution although the contribution of the Bethe-Heitler process 
is non negligible.
An important contamination to DVCS elastic candidates is due to the 
DVCS process with proton dissociation:
\begin{equation}
e^+ + p \rightarrow e^+ + \photon + Y ,
\label{eq:reac2}
\end{equation}
when the decay products of the baryonic system $Y$ are not detected in 
the forward detectors. 
The sum of non elastic DVCS and BH contributions has been estimated to
be $16\pm 8\,\%$ of the final sample~\cite{rainer}, based on the
fraction of events with proton dissociation tagged by the forward
detectors and the detection efficiency of the forward detectors for
proton dissociation events using the DIFFVM Monte Carlo.
The other sources of background to be considered are due to diffractive 
\om\ and \ph\ production, with decay modes to final states including
photons (directly or from $\pi^0$ decay) or $K^0_L$.
The main contributions originate from 
$\omega \rightarrow \pi^0 \gamma$ and $\phi \rightarrow K^0_L K^0_S$  
followed by $K^0_S \rightarrow \pi^0 \pi^0$.
The background arising from $\pi^0$ production, with the decay 
photons reconstructed in a single cluster, in low multiplicity DIS, 
is estimated from data. 
According to the MC simulation the two decay photons are recognised,
in this energy range, in $20\,\%$ of the events in separate clusters. 
The selection has thus been extended to events with at least two
clusters in the LAr calorimeter of energies exceeding 
$0.5\,{\rm GeV}$. 
The two photon invariant mass spectrum is in agreement with 
expectation from diffractive $\omega$ and $\phi$ production.
Background from $\pi^0$ production is therefore inferred to be 
negligible.
Figure~\ref{fig:cont} shows data distributions in 
comparison to the sum of the predictions according to the FFS
calculation and the diffractive \om\ and \ph\ backgrounds.
The BH contribution in the FFS prediction and the \om\ and \ph\
backgrounds are normalised to luminosity.
The DVCS contribution in the FFS prediction is here normalised such
that the sum of all contributions is equal to the total number
of events in the data.
The pure Bethe-Heitler contribution is also shown.
The DVCS signal exhibits different kinematic distributions from the 
Bethe-Heitler contributions, in particular in the polar angle of the 
LAr cluster (Fig.~\ref{fig:cont}c-d) and in coplanarity
(Fig.~\ref{fig:cont}e-f), which is defined as the difference of the 
azimuthal angles of the two clusters and is related to the $p_t$-balance 
of the positron-photon system.
The coplanarity distribution is found to be broader (${\rm rms}=5.3^{\circ}$) 
for the sum of the contribution in the enriched DVCS sample than for the 
Bethe-Heitler dominated control sample (${\rm rms}=3.2^{\circ}$). 
This is attributed to the electromagnetic nature of the BH process 
which implies a steeper $t$ distribution than the DVCS signal.

\section{Cross section measurement and model comparison} \label{sect:cross}

To extract the cross section, the data of the enriched DVCS sample 
have been corrected for detector effects, acceptance and for initial 
state radiation of real photons from the positron line using the Monte 
Carlo simulation.
The bin size has been chosen according to the statistical accuracy and 
is large with respect to resolutions in $Q^2$ ($12\,\%$) and $W$ ($6\,\%$).
The contamination of inelastic BH and DVCS events with proton 
dissociation ($16\pm8\,\%$) is subtracted statistically.
The background contributions from diffractive $\omega$ and $\phi$ 
production, estimated to be $3.5\,\%$ on average and below $6\,\%$ in all bins, 
have been subtracted bin by bin.

\begin{figure}
 \begin{center}
  \epsfig{figure=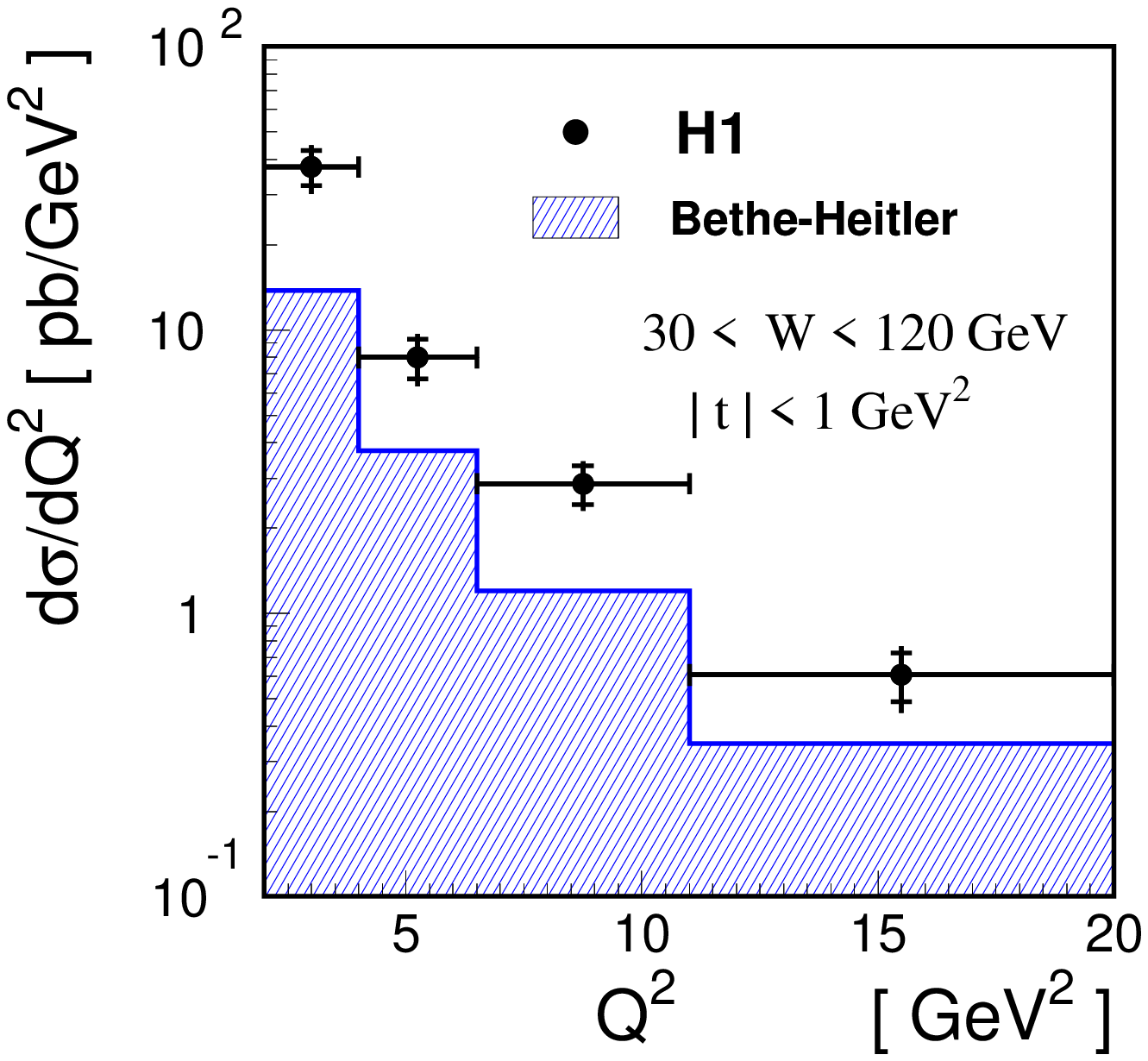,width=0.48\textwidth}
  \epsfig{figure=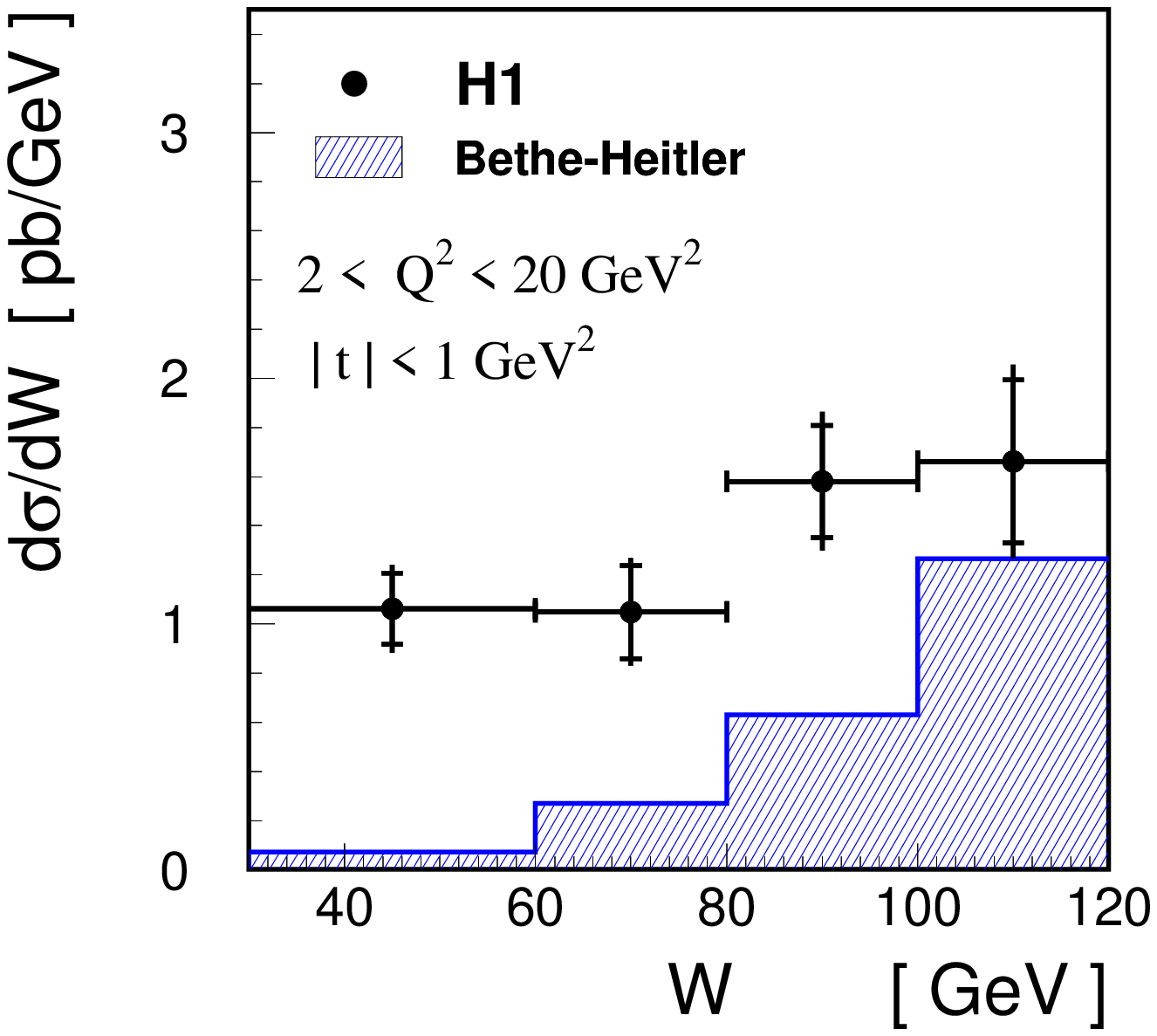,width=0.48\textwidth}
 \end{center}
 \begin{picture}(0,0)
  \put(72,80){\bf a)}
  \put(149,80){\bf b)}
 \end{picture}
 \vspace*{-1.5cm}
 \caption{\sl Differential cross section measurements for the reaction 
  $e^+ p \rightarrow e^+ \gamma p$ as a function of
  $Q^2$ (a) and $W$ (b). The inner error bars are statistical and
the full error bars include the systematic errors added in quadrature.
The hatched histogram shows the contribution of the Bethe-Heitler 
process to the reaction, where, however, $Q^2$ and $W$
do not correspond to the photon
virtuality and the $\gamma^*p$ center of mass energy, respectively.}

 \label{fig:sigep}
\end{figure}

The $e^+p\rightarrow e^+\gamma p$ cross section is presented in
Fig.~\ref{fig:sigep} and in Table~\ref{tab:sigep} differentially 
in $Q^2$ and $W$.
The data are compared to the estimate of the pure Bethe-Heitler 
contribution. 
The total cross section is dominated by the DVCS process at small $W$
values and the Bethe-Heitler process at large $W$ values.
With the present precision, the $Q^2$ slopes of the two processes
appear similar. 
The limited resolution and statistics do not allow the cross section 
measurement differentially in $t$ and the extraction of the slope
from the present data.
The main contribution, of $8\,\%$, to the systematic error arises from 
detector effects and is due to the uncertainty on the measurement of 
the angle of the scattered positron.
Other detector related errors are estimated to be around or below $2\,\%$. 
The second largest systematic error arises from the estimate of the 
contamination of non elastic BH and DVCS ($8\,\%$). 
The total systematic error is found to be around $15\,\%$.

In the leading twist approximation the contribution of the 
interference term to the cross section is proportional to the cosine 
of the photon azimuthal angle\footnote{The photon azimuthal angle is 
defined as the angle between the plane formed by the incoming and 
scattered positrons and the $\gamma^*$ proton plane.}.
Since the present measurement is integrated over this angle, 
the overall contribution of the interference term is negligible.
Therefore the Bethe-Heitler cross section can be subtracted from
the total cross section in order to obtain the DVCS cross section.
The DVCS $e^+p\rightarrow e^+\gamma p$ cross section is then converted 
to a DVCS $\gamma^* p \rightarrow \gamma p$ cross section using the 
equivalent photon approximation (as in~\cite{h1_rho}). 

\begin{figure}
 \begin{center}
  \epsfig{figure=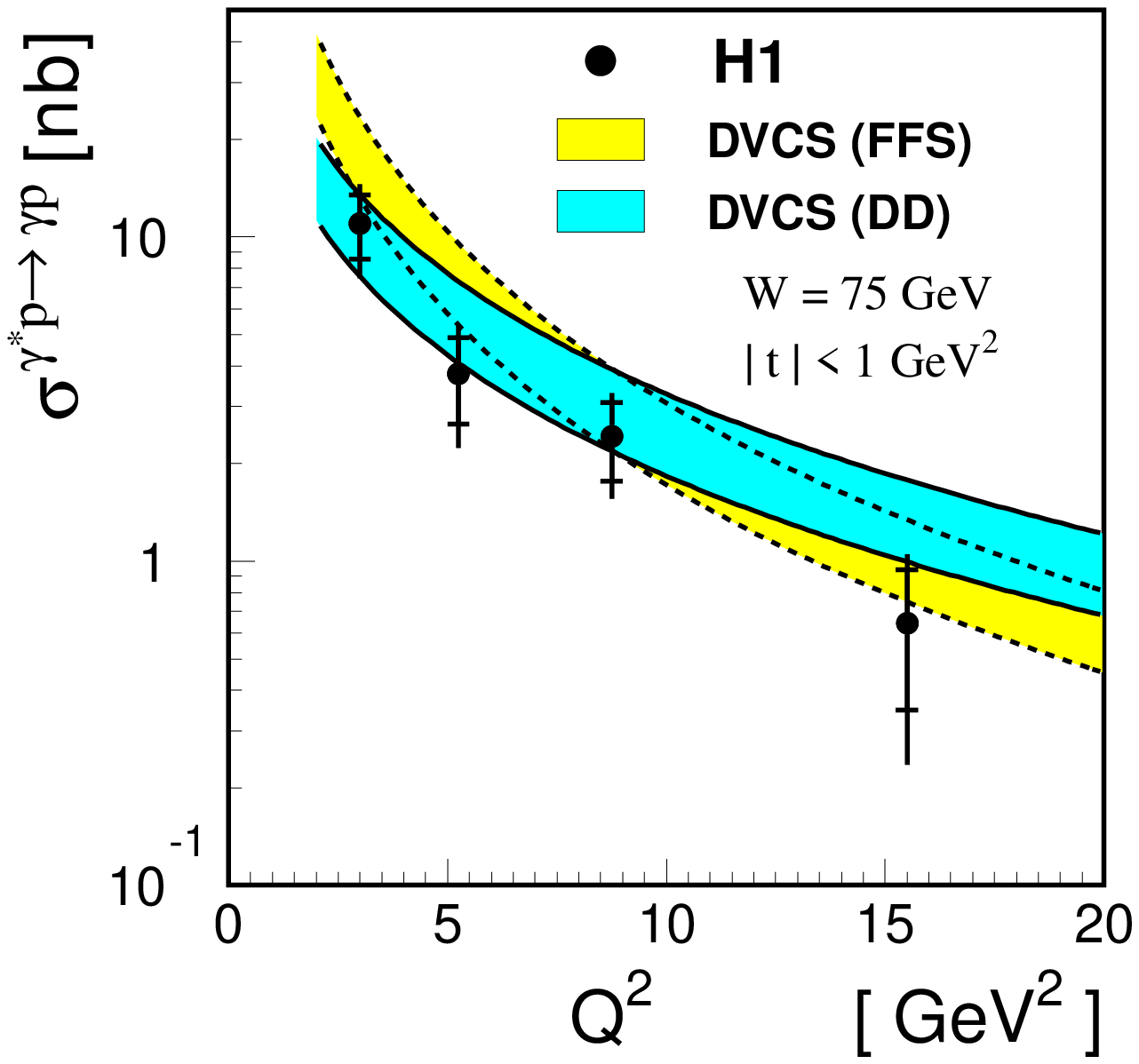,width=0.48\textwidth}
  \epsfig{figure=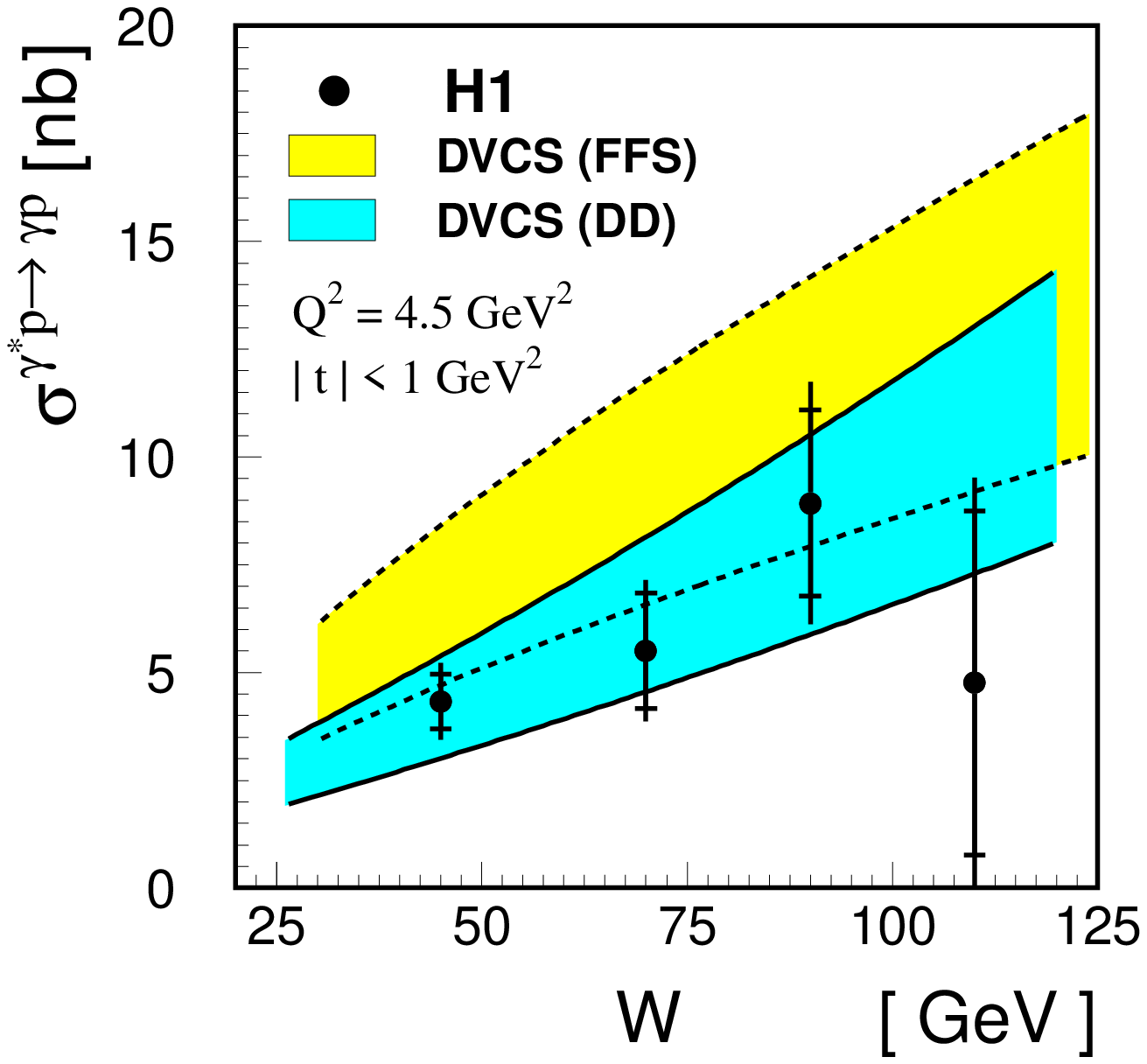,width=0.48\textwidth}
 \end{center}
 \begin{picture}(0,0)
  \put(71,83){\bf a)}
  \put(148,83){\bf b)}
 \end{picture}
 \vspace*{-1.5cm}
 \caption{\sl Cross section measurements for the  
  $\gamma^* p \rightarrow  \gamma p$ DVCS process 
  as a function of $Q^2$ (a) and $W$ (b). 
  The data are compared to the theoretical predictions of 
  FFS~{\rm \cite{Frankfurt:1998at}} and DD{\rm ~\cite{Donnachie:2000px}}.
  The band associated to each prediction corresponds to a variation of
  the assumed $t$-slope from $5 \ {\rm GeV}^{-2}$ (upper bound) to
  $9 \ {\rm GeV}^{-2}$ (lower bound).
  The inner error bars are statistical and the full error bars include 
  the systematic errors added in quadrature.}
 \label{fig:siggp}
\end{figure}

\begin{table}[htbp]
\centering
\begin{tabular}{clllcclll}
\cline{1-4} \cline{6-9}\\[-10.0pt]
  $Q^2$ $\left[{\rm GeV}^2\right]$  & 
  \multicolumn{3}{c}{$d\sigma^{e^+ p\rightarrow e^+ p \gamma}/dQ^2$ $\left[{\rm pb/GeV}^2\right]$} &
\hspace{0.5 cm} & $W$ $\left[{\rm GeV}\right]$  & 
  \multicolumn{3}{c}{$d\sigma^{e^+ p\rightarrow e^+ p \gamma}/dW$ $\left[{\rm pb/GeV}\right]$} \\[1.5pt]
\cline{1-4} \cline{6-9}\\[-10.0pt]
$[2.0,4.0]$    & $\; 37.6 $  & $\pm \, 5.3  $  & $\pm \, 5.1 $ & & 
  $[30, 60]$  & $\: 1.06 $  & $\pm \, 0.15 $  & $\pm \, 0.10$ \\
$[4.0,6.5]$    &  $\; 8.0 $  & $\pm \, 1.3  $  & $\pm \, 1.1 $ & &
  $[60, 80]$  & $\: 1.05 $  & $\pm \, 0.19 $  & $\pm \, 0.11$ \\
$[6.5,11.0]$   & $\; 2.87 $  & $\pm \, 0.46 $  & $\pm \, 0.35$ & &
  $[80,100]$  & $\: 1.58 $  & $\pm \, 0.23 $  & $\pm \, 0.16$ \\
$[11.0,20.0]$  & $\; 0.61 $  & $\pm \, 0.12 $  & $\pm \, 0.11$ & &
  $[100,120]$ & $\: 1.66 $  & $\pm \, 0.33 $  & $\pm \, 0.21$ \\[1.5pt]
\cline{1-4} \cline{6-9}
\end{tabular}
\caption{\sl Differential cross sections for the reaction:
   $e^+ p \rightarrow e^+ p \gamma$, as a function of $Q^2$ and $W$,
in the kinematic domain 
$2 < Q^2 < 20\,{\rm GeV}^2 $, $30 < W < 120\,{\rm GeV}$ and 
$ |t| < 1\,{\rm GeV}^2$.
The first errors are statistical, the second systematic.}
\label{tab:sigep}
\end{table}

The $\gamma^* p$ cross section for the DVCS process is shown in 
Fig.~\ref{fig:siggp} and given in Table~\ref{tab:siggp} as a function 
of $Q^2$ for $W=75\,$GeV, and as a function of $W$ for $Q^2=4.5\,$GeV$^2$.
The systematic errors on the $\gamma^* p$ cross section are due to 
the propagation of the systematic errors on the $e^+p$ cross section 
combined with the bin center corrections error ($7\,\%$).
The data are compared with the predictions by FFS and DD.
The shape of the data is well described by both calculations both
in $Q^2$ and $W$.
The absence of predictions for the $t$-slope leaves an uncertainty on the 
normalisation of the theoretical models.
The band associated to each prediction corresponds to a variation of
the $t$-slope of $5 < b < 9 \ {\rm GeV}^{-2}$, covering the measured
range for light vector meson production~\cite{h1_rho, h1_phi_96}.
Both predictions are consistent with the data within this uncertainty.
It is noted that these data provide constraints also for recent NLO
calculations, which invoke skewed parton 
distributions~\cite{Freund:2001hm}.

\begin{table}[htbp]
\centering
\begin{tabular}{cllllclll}
\cline{1-4} \cline{6-9}\\[-10.0pt]
  $Q^2$ $\left[{\rm GeV}^2 \right]$  & 
  \multicolumn{3}{c}{$\sigma^{\gamma^* p\rightarrow \gamma p}$ $\left[{\rm nb}\right]$} &
\hspace{0.5 cm} & $W$ $\left[{\rm GeV}\right]$  & 
  \multicolumn{3}{c}{$\sigma^{\gamma^* p\rightarrow \gamma p}$ $\left[{\rm nb}\right]$} \\[1.5pt]
\cline{1-4} \cline{6-9}\\[-10.0pt]
 $3.0$  & $11.0$ &$\pm 2.4 $&$\pm 2.5  $ & & $ 45$ & $4.33$ &$\pm 0.64$&$\pm 0.54 $ \\
 $5.25$ & $3.8$  &$\pm 1.1 $&$\pm 1.0  $ & & $ 70$ & $5.51$ &$\pm 1.34$&$\pm 0.86 $ \\
 $8.75$ & $2.43$ &$\pm 0.66$&$\pm 0.54 $ & & $ 90$ & $8.9$  &$\pm 2.2 $&$\pm 1.7  $ \\
$15.50$ & $0.64$ &$\pm 0.30$&$\pm 0.28 $ & & $110$ & $4.8$  &$\pm 4.0 $&$\pm 2.6  $ \\[1.5pt]
\cline{1-4} \cline{6-9}
\end{tabular}
\caption{\sl Measured cross section for the elastic DVCS process 
 $\gamma^* p \rightarrow  \gamma p$ as a function of $Q^2$ for 
 $W=75\,{\rm GeV}$ and as a function of $W$ for $Q^2=4.5\,{\rm GeV}^2$ , both 
 for $ |t| < 1\,{\rm GeV}^2$.
 The first errors are statistical, the second systematic.}
\label{tab:siggp}
\end{table}


\section{Conclusion}

The DVCS process has been studied in the kinematic region 
$2<\, Q^2 <\, 20\,$GeV$^2$, $30<\, W <\, 120\,$GeV and $|t|<\, 1\,$GeV$^2$ 
using a data sample taken by the H1 detector and corresponding to an
integrated luminosity of $8\,$\pbinv.
The cross section for the reaction $e^+p\rightarrow e^+\gamma p$ 
has been measured for the first time and presented differentially 
in $Q^2$ and $W$. 
The DVCS process is observed to dominate over the Bethe-Heitler
process for $W \lsim \, 100$ GeV.  
The $\gamma^* p$ DVCS cross section has been extracted and compared 
to the QCD based predictions~\cite{Frankfurt:1998at, Donnachie:2000px}
which both describe 
the measured $Q^2$ and $W$ distributions within errors.

\section*{Acknowledgements}

We are grateful to the HERA machine group whose outstanding
efforts have made and continue to make this experiment possible. 
We thank the engineers and technicians for their work in constructing 
and now maintaining the H1 detector, our funding agencies for financial 
support, the DESY technical staff for continual assistance and the 
DESY directorate for the hospitality which they extend to the non DESY 
members of the collaboration.
We are grateful to M. Diehl and A. Freund for valuable
discussions.


\end{document}